\documentclass{aastex61}

\newcommand\aastex{AAS\TeX}

\received{}
\revised{}
\accepted{}
\submitjournal{AJ}

\shorttitle{\aastex\ ALMA view of NGC~3557}
\shortauthors{Vila-Vilaro et al.}

\begin{document}

\title{ALMA Observations of the Molecular Gas in the Elliptical Galaxy NGC~3557}

\correspondingauthor{Baltasar Vila-Vilaro}
\email{bvilavil@alma.cl}

\author{Baltasar Vila-Vilaro}
\affil{Joint ALMA Observatory \\
Alonso de Cordova 3107, Vitacura \\
Santiago 763-0355, Chile}

\author{Daniel Espada}
\affiliation{National Astronomical Observatory of Japan \\
2-21-1 Osawa, Mitaka  \\
Tokyo 181-8588, Japan}
\affiliation{SOKENDAI (The Graduate University for Advanced Studies) \\
2-21-1 Osawa, Mitaka \\
Tokyo 181-8588, Japan}

\author{Paulo Cortes}
\affiliation{Joint ALMA Observatory \\
Alonso de Cordova 3107, Vitacura \\
Santiago 763-0355, Chile}

\author{Stephane Leon}
\affiliation{Joint ALMA Observatory \\
Alonso de Cordova 3107, Vitacura \\
Santiago 763-0355, Chile}

\author{Emanuela Pompei}
\affiliation{ESO-Chile \\
Alonso de Cordova 3107, Vitacura \\
Santiago, Chile }

\author{Jordi Cepa}
\affiliation{Instituto de Astrof\'isica de Canarias (IAC) \\
E-38200 La Laguna, Tenerife, Spain}
\affiliation{Departamento de Astrof\'isica, Universidad de La Laguna \\
 E-38206 La Laguna, Tenerife, Spain}


\begin{abstract}

We present the results of CO interferometric observations of the southern elliptical galaxy NGC~3557 with ALMA. We have detected both the CO(1-0) emission line and a relatively strong continuum at 3mm. The continuum shows a flat-spectrum central unresolved source (at our angular resolution of 0$\farcs$7) and two jets, associated with the larger scale emission observed at lower frequencies. The molecular gas in NGC~3557 appears to be concentrated within 250 pc of the center, and shows evidence of organized rotation along the same axis as the stellar component and the symmetry axis of the nuclear dust absorption reported in the literature. We obtained M$_{H_2}$=(9.0$\pm$2.0)x10$^7$ M$_\odot$ of molecular gas, which has an average CO(2-1) to CO(1-0) line ratio of 0.7, which is relatively high when compared with the values reported in the literature for bona-fide ellipticals observed with single-dish telescopes. NGC~3557 shows further a high excitation peak (i.e., CO(2-1)/CO(1-0) $\approx$ 1.1$\pm$0.3 offset 0$\farcs$7 from the center, which appears to be associated with a region of higher velocity dispersion that does not share the overall rotation pattern of the molecular gas, but aligned with the radio jet. The molecular gas disk in this object appears to be stable to local gravitational instabilities.  

\end{abstract}




\section{Introduction} \label{sec:intro}

Several single-dish surveys of CO(1-0) and CO(2-1) emission in early-type galaxies have been reported
in the literature (e.g., \citealp{dav14,wel10,oca10,com07,sag07,kna96,wik95}, etc.). Some of the brightest objects (in CO) have
been further mapped with mm-wave interferometers at resolutions of a few arcsec (e.g., \citealp{tem18,boi17,esp12,cro11,you02}). Very few studies have been carried out on the physical
conditions of the molecular gas in these objects (e.g., \citealp{bay13,bvv03}). From all these studies one is led to conclude that, in terms of bulk molecular gas properties, early-type galaxies contain significantly less molecular gas than spiral galaxies of comparable total mass, and that the molecular gas is sub-thermally excited. Regarding the overall structure and distribution of the molecular ISM, it appears to be concentrated within 1/10 of the optical diameter of the hosts, and in some cases, distributed in disk-like structures with organized rotation (e.g., \citealp{dav13,you02}). Furthermore, the star formation efficiencies are comparable or lower to those of spirals, suggesting similar ongoing star formation processes (e.g., \citealp{dav14,wik95}). One important caveat on the conclusions summarized above is that, most of the surveys mentioned contain a very high fraction of lenticular and early-type spirals. Furthermore, most of the true ellipticals are non-detections in these samples. This implies that the conclusions drawn from these surveys cannot be easily extrapolated to bona-fide ellipticals. 

The evidence for star formation in genuine elliptical galaxies has been the object of a long-standing discussion, mainly because there are several mechanisms that can explain some of the traditional star formation tracers in these objects {\it without} requiring the presence of ongoing star formation (Huang and Gu 2009). For some objects, the combination of several of these tracers observed simultaneously in a host (i.e., H$\alpha$ emission, presence of significant amounts of young stars, local warm IR colors, presence of local free-free emission, etc.) have been invoked as proof of star formation (see for instance, Young 2002, and Cepa et al. 1998). This poses several questions regarding the dynamical structures present in the molecular ISM of elliptical galaxies, and on the possible mechanisms that trigger star formation in these objects as compared with those in spirals and lenticular galaxies. In particular: (1) whether star formation is a locally-triggered process or it involves large-scale gravitational structures like “spiral arms”/bars; (2) whether there are  molecular gas structures similar to Giant Molecular Clouds (hereafter GMCs) and/or Giant Molecular Associations (hereafter GMAs) acting as parents of the star
formation. Most of the published interferometric observations of molecular gas in elliptical galaxies (Crocker et al. 2011, 2012, Young 2002, Cepa et al 1998) have not been of enough sensitivity and spatial resolution to address most of the issues above (with the exception of the recent paper by \citet{boi17}, which contains three objects unequivocally classified as elliptical galaxies), and most of the objects observed are non-typical, in the sense that they are quite CO-bright, and contain significantly more molecular gas mass than the average. The main reason for the dearth of observations, apart from the selection effects, is the peculiar combination of requirements for such a project. That is, the overall weakness of the molecular emission in true ellipticals, combined with small apparent size of the structures to map (both due to intrinsic size and the fact that there are not many nearby elliptical galaxies). To be able to answer some of these questions, it was therefore decided to carry out a survey using the high-sensitivity ALMA interferometer, and we address in this paper the results for one of our targets.
Some of the nearby elliptical galaxies are also radio galaxies, with well defined large scale jets. The orientation of these jets with respect to the molecular gas structures that might exist in these objects is also important, because the radio plasma could interact with the molecular gas, either triggering star-formation or quenching it by local heating \citep{fab12}. Therefore, interferometric studies of the molecular gas structures present in nearby radio galaxies can provide a deeper insight into the triggering of star formation at cosmological distances, where detailed studies of the role of the radio jets are more difficult due to the poorer spatial resolutions involved.     

NGC~3557 is a bona-fide southern-sky elliptical galaxy (E3) at a distance of 40 Mpc (1$\arcsec$ is 198 pc at this distance\footnote{For distace derivations, we have assumed a cosmology model with H$_0$=73~km~s$^{-1}$Mpc$^{-1}$, $\Omega_{matter}$=0.27, and $\Omega_{vacuum}$=0.73}), and member of a small group of galaxies \citep{bro06}. It has been classified as a LINER (e.g., \citealt{ann10}) and as a flat-spectrum radio galaxy \citep{hea07}, with a jet that bends at distances of a few arcmin from the center \citep{sch02}. Detections of several of the ISM components of this object have been reported in the literature. Dust has been observed {\it both} as FIR emission \citep{pas09} and as absorption against the central stellar continuum \citep{lau05}. Nuclear optical line emission has been reported from spectroscopy (\citealp{ver86,ram05}) as well as narrow-band photometry \citep{gou94}. The MIR spectrum of NGC 3557 is of the most common Class-2 type, which is currently associated with a post-star-formation phase \citep{veg10}. Regarding the cold gas component, atomic gas (HI) has been reported as non-detection in several works \citep{ser10}, while molecular gas emission has been detected in single-dish observations \citep{pra10}. NGC~3557 was included in our sample because of its relative proximity and CO(2-1) brightness (as reported by Prandoni et al. 2010), which suggested the possibility to study the molecular structures in detail. Since the object is considered to be in a stage where little current star formation may be happening, it was considered important as a representative of the molecular gas structures to be expected in post-star formation scenarios. Furthermore, the disk-like nuclear dust distribution seen in the photometry by \citet{lau05} indicated a possible presence of a molecular gas disk, which is a very interesting dynamical structure linked to the presence of organized angular momentum at those spatial scales. 

\section{Observations and Data Reduction} \label{sec:obs}

As part of the ALMA programme 2015.1.00591.S (P.I.: Baltasar Vila-Vilar\'o), we observed the southern elliptical galaxy NGC~3557 in the $^{12}$CO(1--0) line (ALMA Band 3). One of the requirements of the programme was to achieve angular resolutions of $\approx$100~pc on all the targets. In the case of NGC~3557, this required the combination of two sets of observations done with different ALMA configurations (see Table \ref{tab:obs}). Given that observations in the literature indicate that the molecular gas in early-type galaxies is usually concentrated within 1/10 of the optical diameter of the host \citep{bvv03}, no ACA 7m-Array observations were required, given the HPBW of the ALMA 12m antennas at 3 mm (i.e.,$\approx$54$\arcsec$) and the 4$\farcm$1 x 3$\farcm$0 optical dimensions of NGC 3557. The Maximum Recoverable Scale (MRS) of our observations were 25$\farcs$2 and 6$\farcs$5 for the configurations C36-2 and C40-5, respectively. 

\begin{deluxetable*}{ccccccc}
\tablenum{1}
\tablecaption{ALMA CO(1--0) Observations\label{tab:obs}}
\tablewidth{0pt}
\tablehead{
\colhead{Start Date (UT)} & \colhead{ALMA Configuration} & \colhead{\# of Antennas} & \colhead{Baseline Lengths (m)} & \colhead{On-Source Time (sec)} & \colhead{PWV (mm)} & \colhead{Average Tsys (K)}
}
\startdata
Dec 26$^{th}$ 2015, 08:14:56 & C36-2 & 34 & 15.1 -- 310.2 & 608 & 3.5 & 90 \\
Jul 23$^{rd}$ 2016, 20:50:38 & C40-5 & 40 & 15.1 -- 1100.0 & 608 & 0.7 & 85 \\
\enddata
\end{deluxetable*}

Four spectral windows were used in the correlator set-up, one centered on the redshifted $^{12}$CO(1--0) line using the Frequency Division Mode (FDM) with 3840 channels covering a 1.875 GHz bandwidth (spectral resolution $\approx$2.5 km~s$^{-1}$), and three continuum spectral windows using the Time Division Mode (TDM) with 128 channels covering 2 GHz bandwidths each. The continuum spectral windows were set to cover the whole 4-8 GHz IF range in the LSB, and one in the USB avoiding the region where the expected $^{12}$CO(1--0) line emission was expected to be. Following standard ALMA practice, the calibration of the phase effects of the highly-variable water vapour content of the atmosphere was achieved using the data of Water Vapour Radiometers installed on each antenna, and the System Temperature measurements were done periodically (every 15 minutes) using two loads located on top of the receiver cryostat. Absolute flux scales were derived from observations of the quasar J1107-4449, which also was used as the bandpass calibrator. As phase calibrator the quasar J1126-3828, which is located just a couple of degrees away from near NGC~3557, was used. 

An initial investigation of the data quality was done on the delivered data products that had been processed by the standard ALMA pipeline (version 4.5.1). We then used the same Common Astronomy Software Applications (CASA) package version to re-calibrate the data in preparation for a data combination of the datasets taken with different configurations. For the continuum spectral windows, the data was merged with the {\it concat} command weighting, in addition to the weights already included in the metadata, for the ratio of the final noise levels of the synthesized images (i.e., parameter {\it visweightscale} in {\it concat}). For the spectral line windows, the additional weighting was done on the ratio of the number of samples in the regions of the uv-plane where both configurations overlapped. To avoid issues with continuum subtraction that can occur if done after data merging, we subtracted the continuum (uvcontsub) in the spectral line windows {\it prior} to merging them. The resulting visibility files were then imaged using the {\it clean} task with a Briggs weighting using a robust parameter of 0.5. For the continuum images, an additional self-calibration step was performed to better study the low-level continuum emission associated to the jets in this target. For the spectral line spectral windows, data cubes were generated with multiple final velocity resolutions (i.e., 5, 10, 20 and 40 km~s$^{-1}$, respectively) to study in detail the different features of the CO(1-0) emission. Maps using {\it only} the more extended configuration were also produced to visualize better possible clumps in the molecular gas. The final synthesized beam HPBW for the CO(1-0) data was 0$\farcs$79$\times$0$\farcs$74 along PA\footnote{Throughout this paper, position angles will be defined following the usual standard, that is origin of angles pointing North and counted towards the East.} 75$\degr$. Additionally, the achieved 3mm continuum sensitivity was 26$\mu$Jy/beam, and for CO(1-0) emission 0.8~mJy/beam (for 20 km~s$^{-1}$ binning), respectively. 

In order to derive the physical conditions of the molecular gas in NGC~3557 (addressed in Section 8 of this paper), the continuum spectral index (Section 3), and to corroborate the molecular gas rotation curve derived from our data (see Section 6 of this paper), we have used the public ALMA data of project 2015.1.01572.S (PI: Prandoni), which consists of CO(2-1) observations of the same target\footnote{For comparison with the data listed in Table 1, the date of the observations was 2016-06-03 23:48:09 UTC and the ALMA array included 37 antennas covering a baseline range of 16.7 -- 650.3 m. The PWV was 1.73 mm, the time on source 1368s and the average system temperatures were 80K, respectively.}. For reference, we include here information on the synthesized beam size and sensitivity of those data. The synthesized beam and orientation of the CO(2-1) were 0$\farcs$59$\times$0$\farcs$54 along PA 85$\degr$, respectively. Additionally, the achieved 1mm continuum sensitivity was 107.6$\mu$Jy/beam, and for CO(2-1) spectral emission 0.45~mJy/beam (for 40 km~s$^{-1}$ binning), respectively.

\section{Continuum Observations and Spectral Index} \label{sec:cont}

\begin{figure}
\plotone{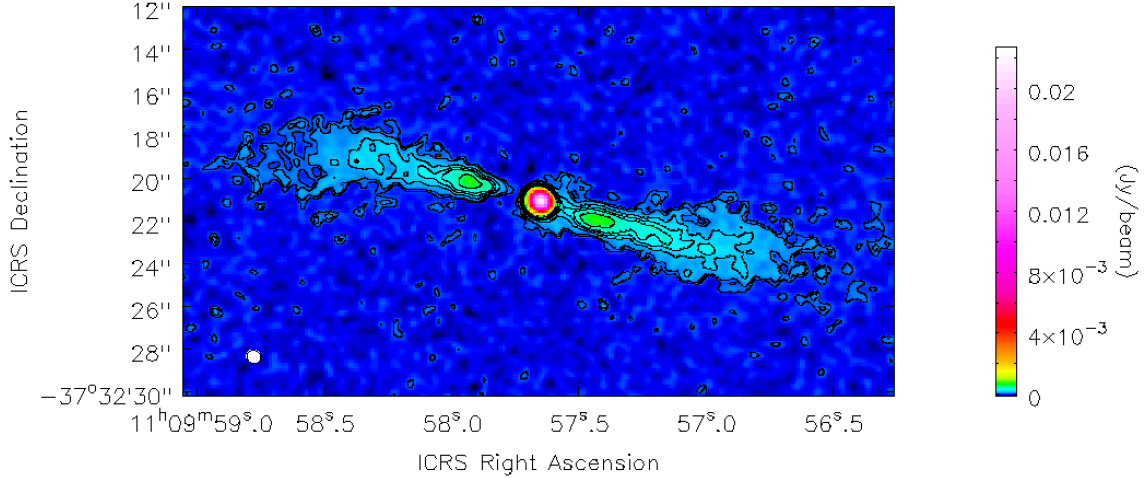}
\caption{Continuum ALMA image of the inner 36$\arcsec$ of NGC~3557 at 3mm. The synthesized beam (0$\farcs$61$\times$0$\farcs$65 along PA 61$\degr$) is shown in the lower left side of the figure. The achieved sensitivity is 26$\mu$Jy/beam, and the contour levels are at (0.52, 0.80, 1.6, 2.4, 3.2, 5.2)~10$^{-4}$ Jy beam$^{-1}$, respectively.\label{fig:fig1}}
\end{figure}

Figure \ref{fig:fig1} shows the continuum emission in the inner 36$\arcsec$ of NGC~3557. The emission is dominated by an unresolved central source and two narrow jets straddling the nuclear source at a PA of $\approx$74$\degr$. A fit to the central unresolved source gives a peak position of ICRS $\alpha$=11:09:57.650 and $\delta$=-37:32:21.07, which is very similar to the coadded 2MASS survey position in the JHK bands, quoted in \citet{skr06} as $\alpha$=11:09:57.639 and $\delta$=-37:32:21.04. We therefore, assume for the rest of this paper that the central unresolved radio continuum source marks the position of the center of NGC~3557.

The jet on the SW of the nucleus appears to be connected with the central unresolved source by a narrow bridge, while the jet of the NE is apparently disconnected from it. Both jets contain bright emission clumps near the center, which resemble an FR I-type radio structure. The jets appear to broaden at radial distances of 5$\arcsec$ (i.e., 990pc) from the unresolved central component. The overall orientation of the jets is in agreement with the larger-scale lower-frequency VLA observations in \citet{sch02}. The jet to the SW appers to be brighter than the jet on the NE, which could suggest the presence of Doppler boosting. We can extract some information on the spectral index of the different components by comparing them with the 1mm continuum emission in the publicly available band 6 ALMA observations (2015.1.01572.S by Prandoni et al.\footnote{These observations achieved an angular resolution of 0$\farcs$5}). As shown in Table \ref{tab:specind}, the band 6 data in the ALMA Archive do not have enough sensitivity (not even after self-calibration) nor do they recover the required angular scales to see the jet (MRS is 3$\farcs$5) to allow us to derive a spectral index for the inner bright clumps in the jets straddling the nucleus a 3-4$\arcsec$ from the center (a typical jet spectral index of -0.7 would imply a decrease of 38\% in the flux for the jet clumps in the band 6 observations compared with our data), but at least we can derive it for the central unresolved source, which, as expected, turns out to be a typical flat-spectrum. For the jet clumps, we can only state that the spectral index is negative. 

\begin{deluxetable*}{ccccc}
\tablenum{2}
\tablecaption{Continuum spectral indices\label{tab:specind}}
\tablewidth{0pt}
\tablehead{
\colhead{Component} & \colhead{Distance to Center/PA} & \colhead{Flux (3mm)} & \colhead{Flux (1mm)} & \colhead{Spectral Index} \\ 
\colhead{} & \colhead{(arsec/degrees)} & \colhead{(mJy)} & \colhead{(mJy)} & \colhead{} 
}
\startdata
Center & 0./0. & 23.7 & 24.1 & 0.02 \\
SW Clump & 2.8/254 & 2.1 & $<$ 1.7 & $<$ -0.3 \\
NE Clump & 3.5/74 & 2.3 & $<$ 1.7 & $<$ -0.4 \\
\enddata 
\tablecomments{The 1mm data of the Prandoni et al. project in this Table were measured on smoothed images that matched the angular resolution of our CO(1-0) data. Furthermore, the areas on the images used in the estimation of the 1mm upper limits were the same as those in our data.}
\end{deluxetable*}

\section{Fast or Slow Rotator} \label{sec:class}

To discern whether NGC~3557 is a true fast rotator, as defined in \citet{ems07} and \citet{ems11}, we have used multiwavelength data available in the literature. The K-band photometry published in \citet{pah99} gives an effective ellipticity of $\epsilon_{e}$ = 0.251 for this object. A value for the normalized angular momentum parameter $\lambda_{Re}$ \citep{ems07} could be derived from the long-slit stellar spectroscopy along the major axis of this galaxy in \citet{bro07}, obtaining $\lambda_{Re}$ $=$ 0.7 for an effective radius of 14$\arcsec$ \citep{pah99}. Looking at Figure 6 (bottom) of \citet{ems11}, it is clear that NGC~3557 sits in the upper region of the fast rotator area, with a separation between slow and fast rotators defined as 0.31~$\sqrt{\epsilon_{e}}$ = 0.16 for the effective ellipticity of 0.251 in \citet{pah99}. The same plot also suggests that the actual value of $\lambda_{Re}$ for the whole galaxy may be smaller than the value reported here (based on major-axis spectroscopy only); for the effective ellipticity of this object, there are no objects in the observed sample of \citet{ems11} that have values above 0.65. We can additionally compare $\epsilon_{e}$ with the ratio of the rotation velocity to the central velocity dispersion (i.e., V/$\sigma$) in NGC3577, which is reported in \citet{bus92}, V/$\sigma$=1.05. Again we obtain, when compared with Figure 6 (bottom) of \citet{ems11}, that this object is in the region of fast rotation. We conclude therefore that NGC~3557 is a fast rotator and we proceed to discuss its properties below.

\section{Molecular Gas Properties} \label{sec:distrib}

\begin{figure}
\gridline{\fig{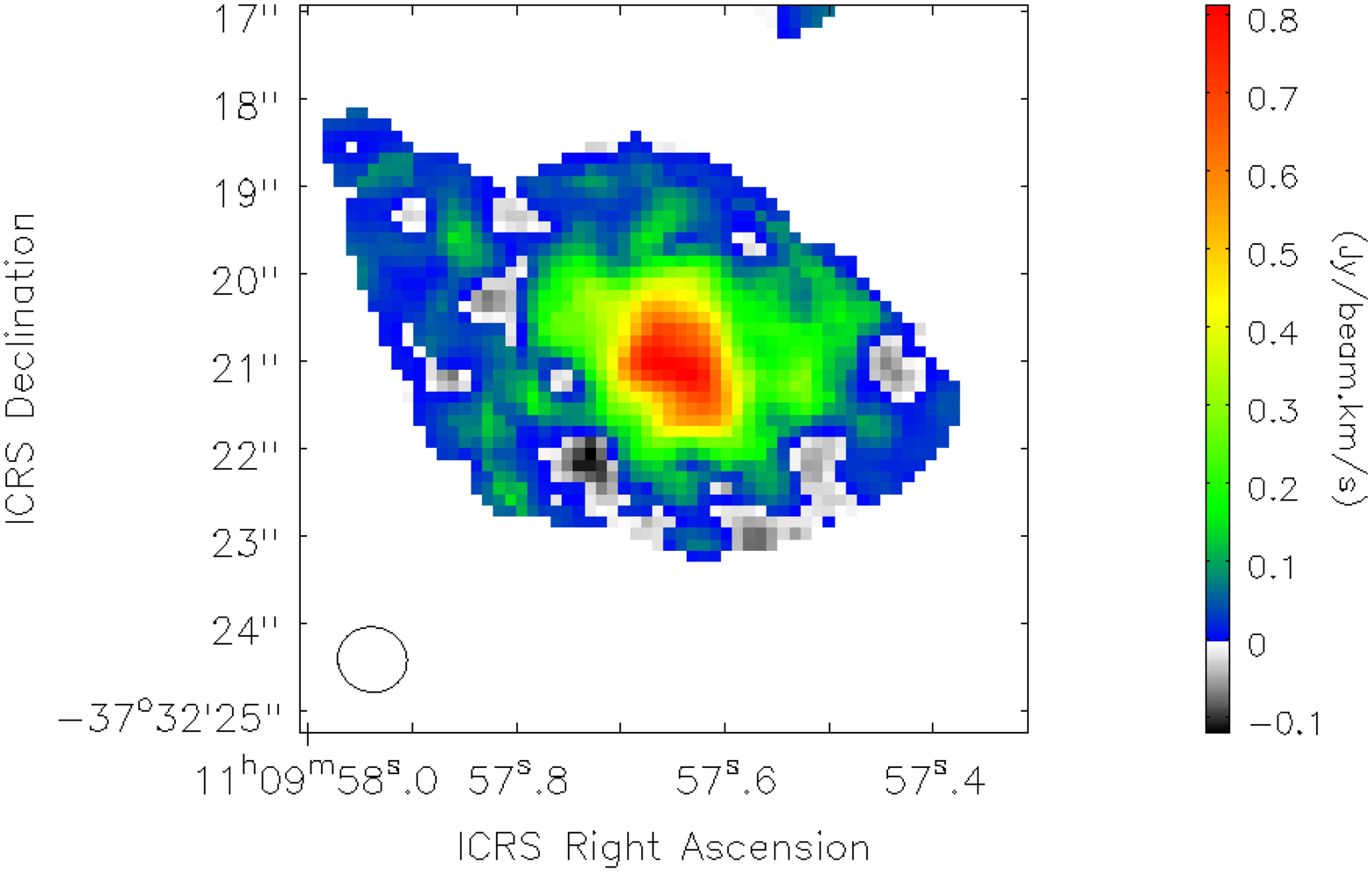}{0.55\textwidth}{(a)}}
\gridline{\fig{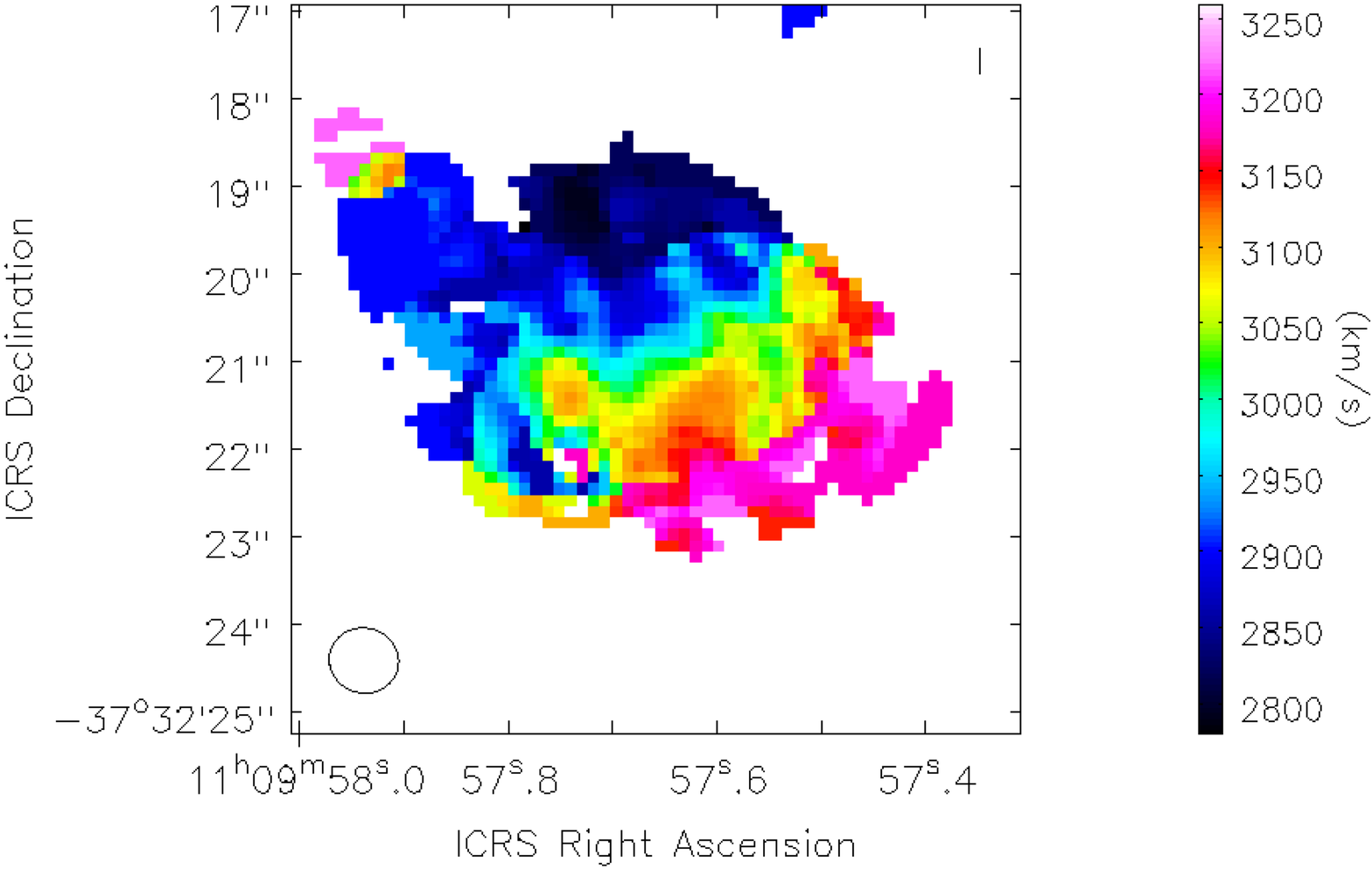}{0.55\textwidth}{(b)}}
\gridline{\fig{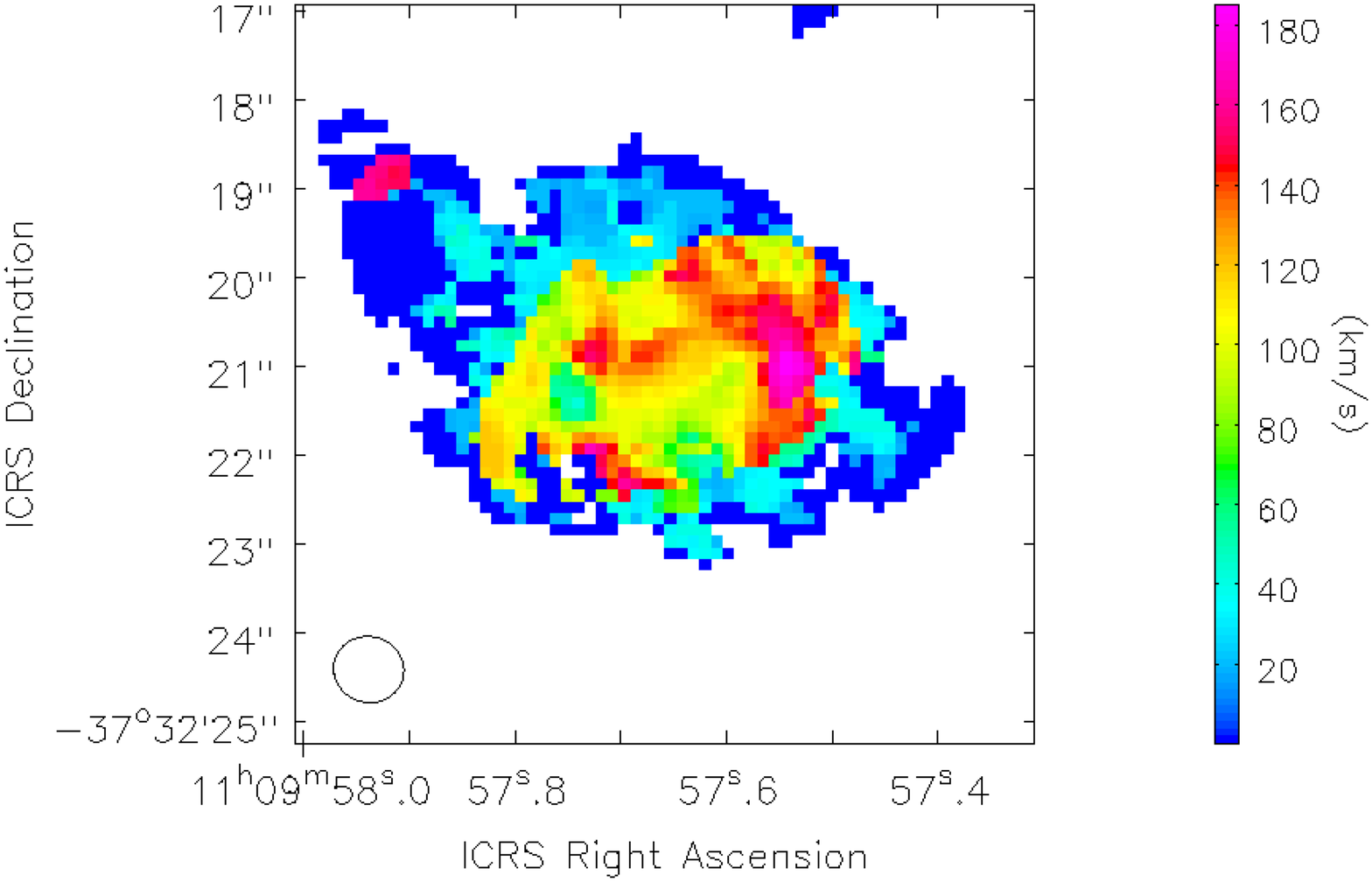}{0.5\textwidth}{(c)}}
\caption{CO(1-0) (a) Integrated intensity, (b) velocity field, and (c) velocity dispersion maps of the molecular gas in the central regions of NGC~3557. The same mask, excluding regions with SNR below 2.5 times the noise RMS on the channel maps\textsuperscript{a}, was applied to all three plots. The synthesized beam is shown on the lower left of the panels.\label{fig:fig2}}
\small\textsuperscript{a} {\bf The regions with low SNR are derived from a smoothed datacube using a smoothing kernel of 20x30x3 pixels. This generates the mask that is then applied to the original datacube.}
\end{figure}

\begin{figure}
\gridline{\fig{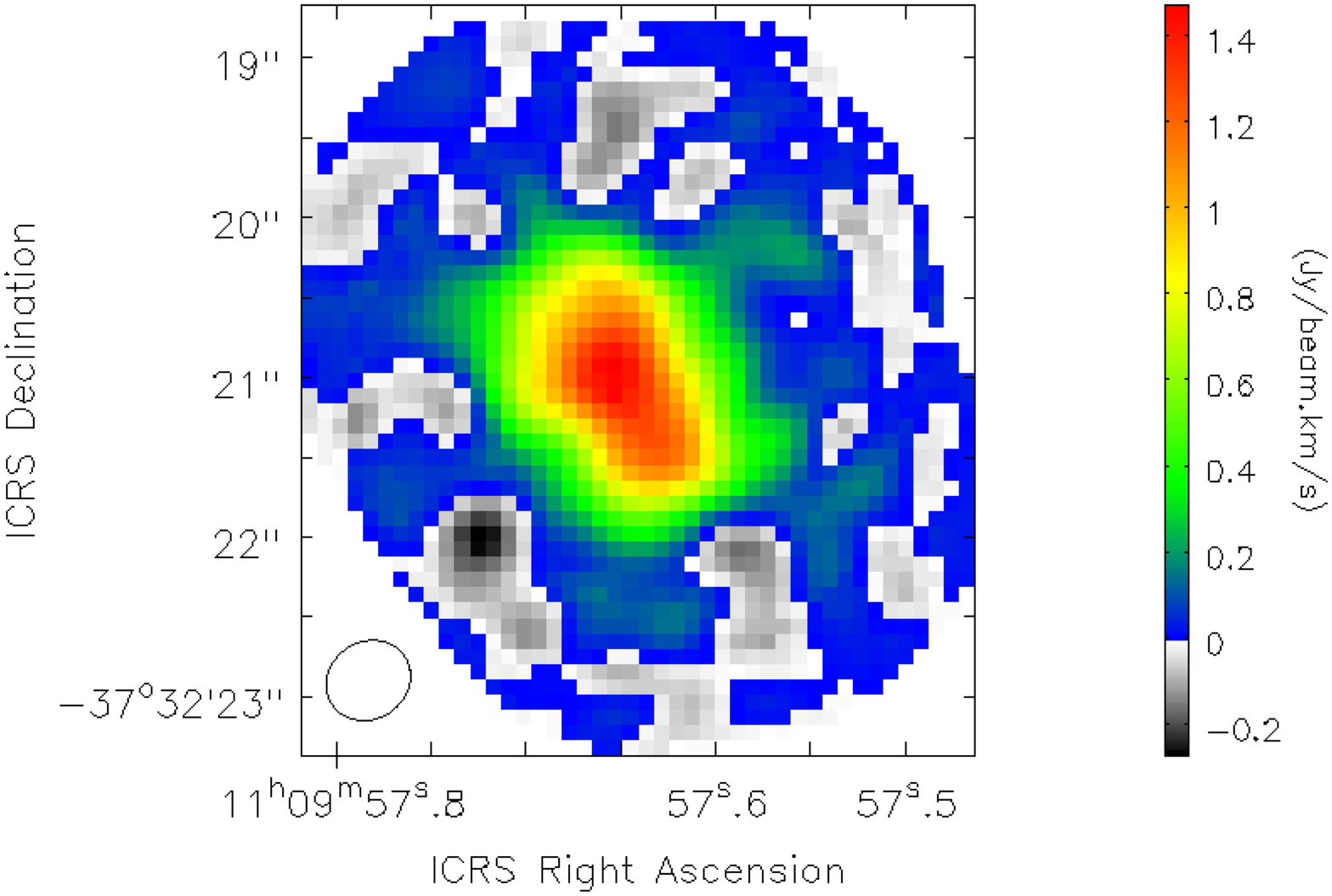}{0.7\textwidth}{(a)}}
\gridline{\fig{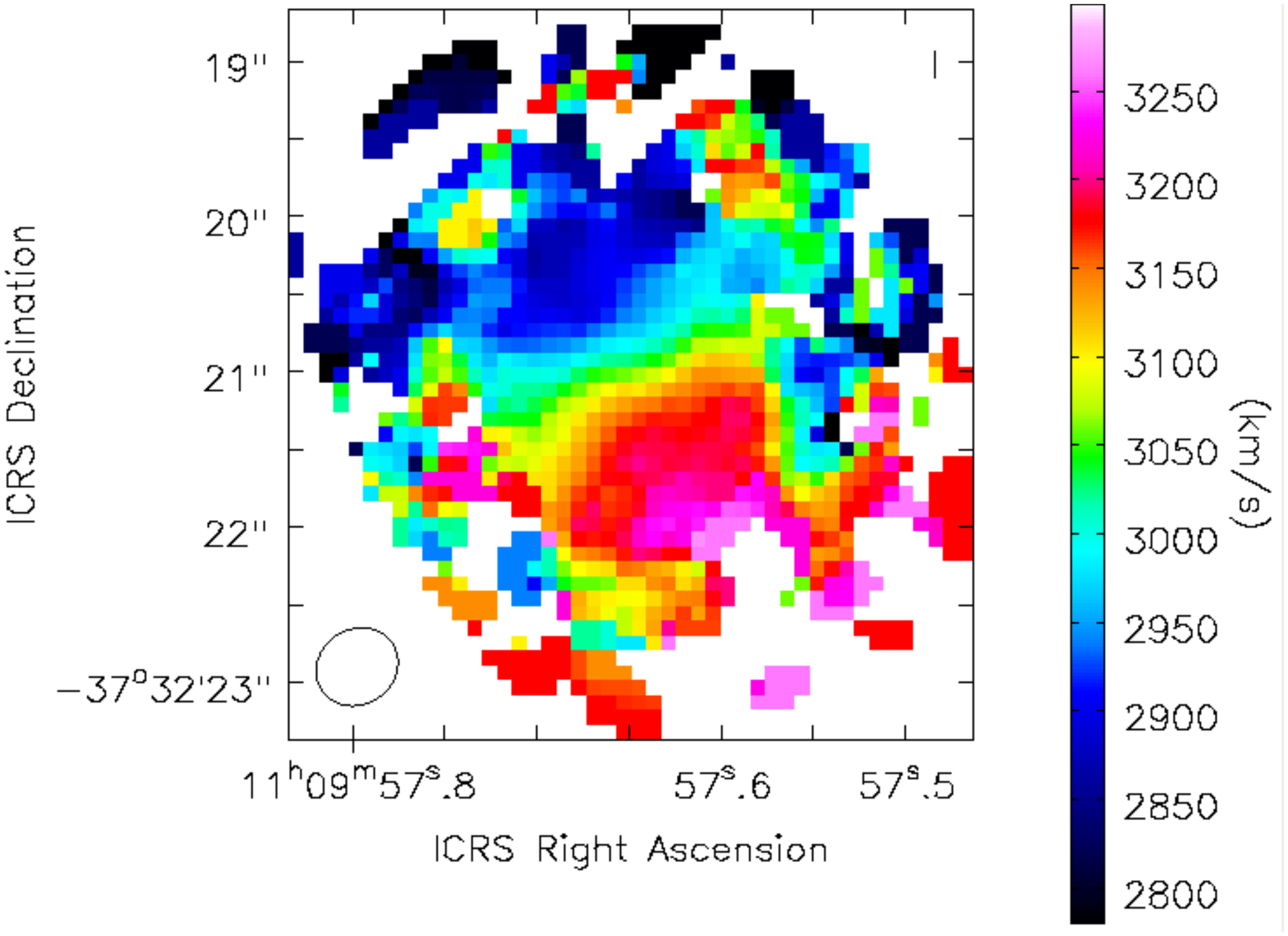}{0.7\textwidth}{(b)}}
\caption{{\bf CO(2-1) (a) integrated intensity, and (b) velocity field  of the molecular gas in the central regions of NGC~3557, covering the same velocity range as that of the CO(1-0) maps in Figure \ref{fig:fig2}. The mask used was also derived following the same procedure as that in Figure \ref{fig:fig2}. The synthesized beam is shown on the lower left.}\label{fig:fig3}}
\end{figure}

The integrated-intensity CO(1-0), velocity field and velocity dispersion maps of NGC~3557 for the velocity range 2740 -- 3220 km~s$^{-1}$ are shown in Figure \ref{fig:fig2}. The CO(1-0) distribution is clearly concentrated around the position of the nucleus, with a maximum radial extension of about 1$\farcs$5--2$\arcsec$ (i.e., 297--396 pc in projected radial distance). Most of the emission appears to be concentrated within a radial distance of 1$\arcsec$ in a structure elongated along PA~=~30$\degr$, while the rest of the emission lies in a lower-luminosity envelope around. The total integrated CO(1-0) flux is 4.5$\pm$0.9 Jy~km~s$^{-1}$, which, using the standard conversion factors\footnote{i.e., X$_{CO}\approx$2.8$\times$10$^{20}$ cm$^{-2}$ (K~km~s$^{-1}$)$^{-1}$} (\citealp{sol87,bol13}) implies a molecular hydrogen mass of M$_{H_2}$=(9.0$\pm$2.0)x10$^7$ M$_\odot$, or a total molecular gas mass including Helium of M$_{mol}$=(12.2$\pm$2.4)x10$^7$ M$_\odot$. These numbers differ quite significantly from those reported in \citet{pra10}, i.e. log(M$_{H_2}$[M$_\odot$]) = 9.02. Since several careful revisions of our data reduction did not reveal any problems, and the values in \citet{pra10} are based on single-dish CO(2-1) observations, it was opted for checking the CO(2-1) APEX telecope data used in their paper (which can be publicly accessed via the ESO Archive), and also the public ALMA NGC~3557 CO(2-1) observations of project 2015.1.01572.S {\bf (Figure \ref{fig:fig3} shows the moments 0 and 1 images, for the same velocity range as in Figure \ref{fig:fig2})}. We find that the data used in \citet{pra10} was very severely affected by a standing-wave component, that if not removed properly, can cause fictitious CO(2-1) emission at the levels reported in that paper. Since this fact is not reported in that paper, we cannot ascertain how this issue was handled. However, we have also found in the same APEX archive datasets by other observers on the same target that (see for instance, datasets for project E-083.B-0543A), when reduced, do not reproduce the CO(2-1) values in \citet{pra10}, giving upper limits for the CO(2-1) emission of 0.4 K~km~s$^{-1}$ compared with the 1.52 K~km~s$^{-1}$ in \citet{pra10}. Furthermore, the ALMA CO(2-1) data gives an integrated flux of 7.5$\pm$0.3 Jy~km~s$^{-1}$, which again can be compared with the 89.7$\pm$11.7 Jy~km~s$^{-1}$ that would be derived from \citet{pra10} using a conversion factor Jy/K of 39 for the APEX antenna. Based on all this evidence, we conclude that our value for the molecular gas mass in NGC~3557 is more plausible, and we will use it for the rest of this paper.

\begin{figure}
\plotone{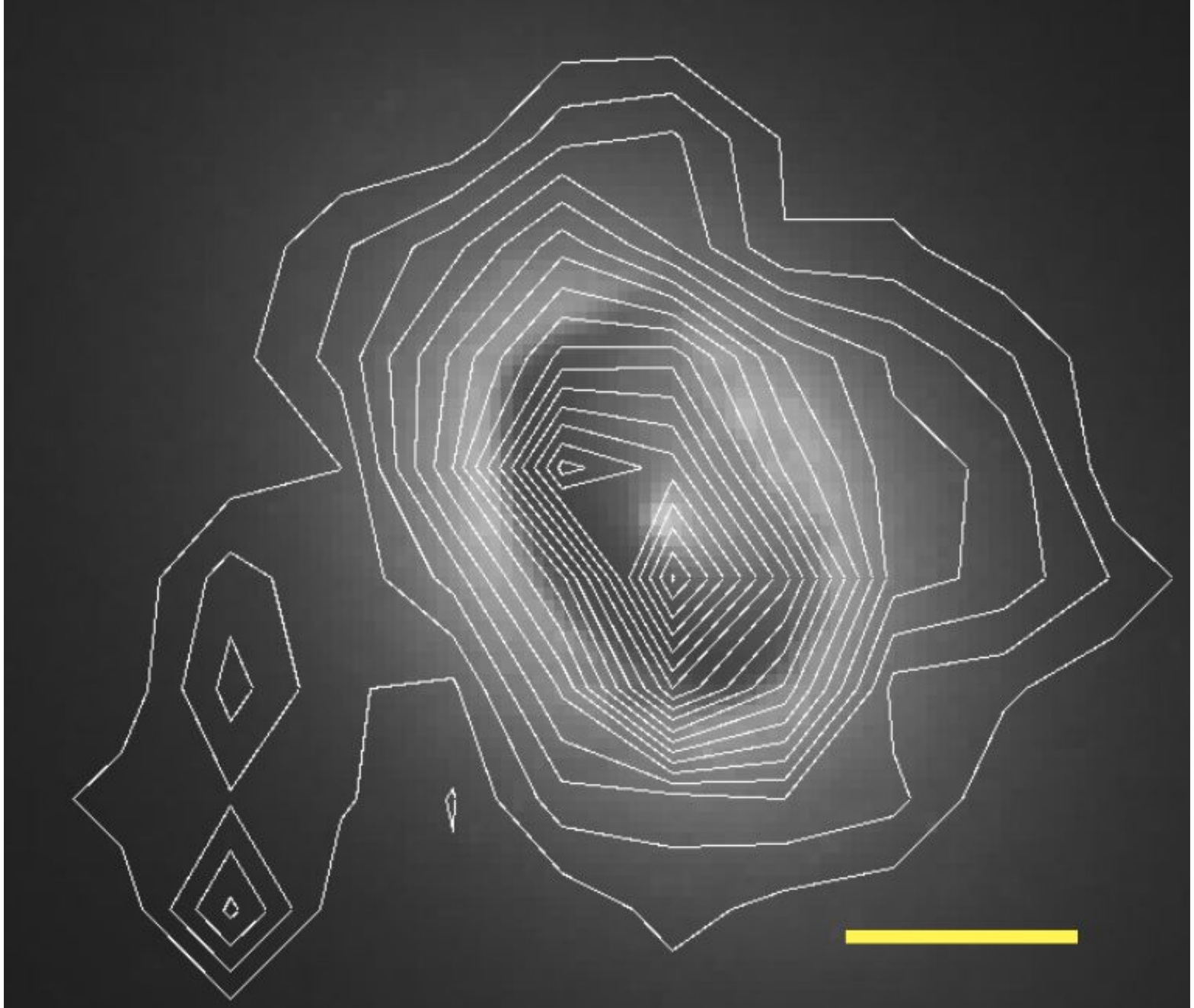}
\caption{Contour map (white lines) of the CO(1-0) integrated intensity superposed on the HST dust absorption map in \citet{lau05}. The agreement between the two suggests that the brightest areas in the CO(1-0) map are associated with the dust disk. The horizontal yellow line covers 1$\arcsec$. \label{fig:fig4}}
\end{figure}

\begin{figure}
\plotone{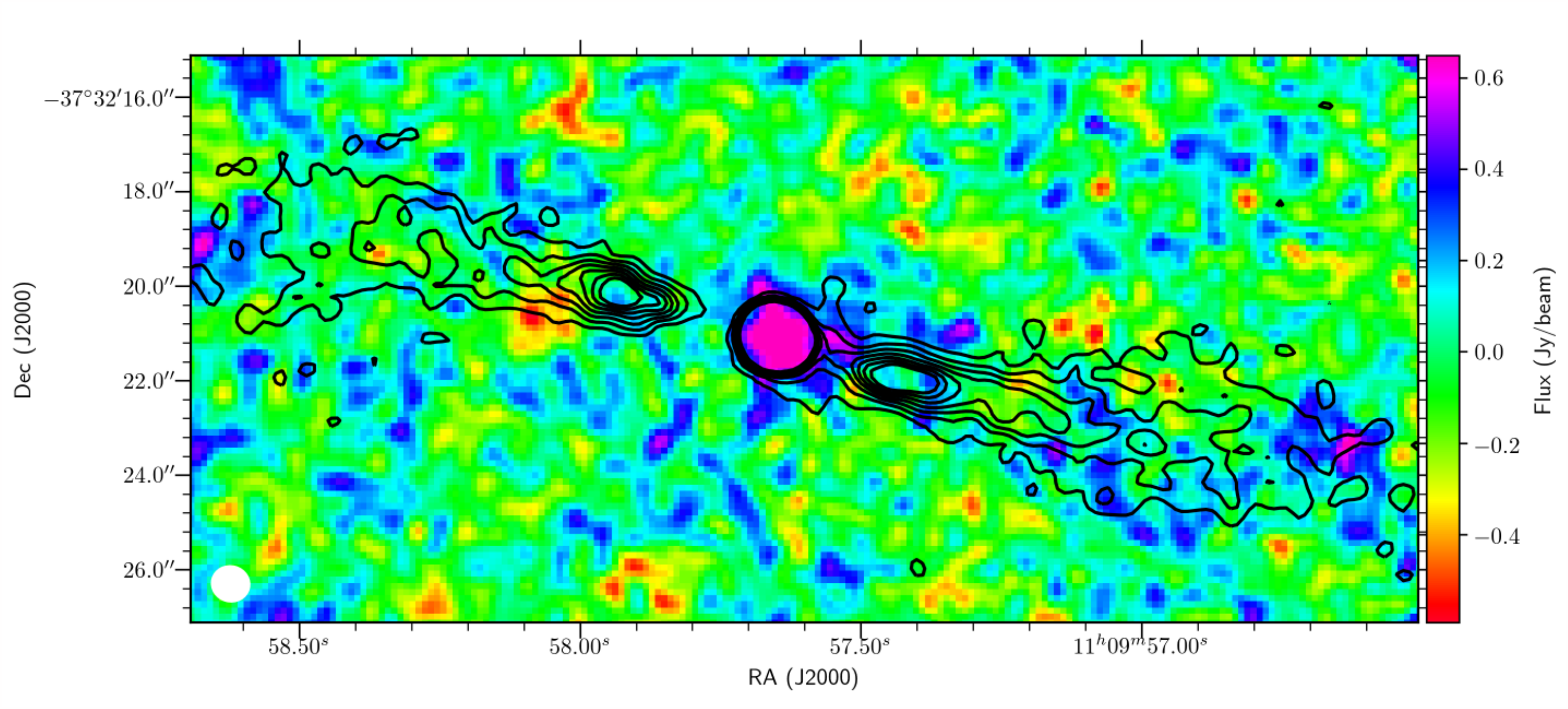}
\caption{Contour map (black lines) of the 3mm continuum image described in Section \ref{sec:cont} superposed on the integrated CO(1-0) line emission. The molecular gas seems to lie mostly around the unresolved central continuum peak.\label{fig:fig5}}
\end{figure}

We have compared the molecular gas distribution with the dust ring-like structure shown in the Hubble Telescope broadband images \citep{lau05}. The superposition of both in Figure \ref{fig:fig4} clearly shows that the inner 1$\arcsec$ region of brighter CO(1-0) emission is very well aligned with the overall dust structure. Furthermore, given the higher obscuration in the eastern side of the dust ring, we assume that that is the side closer to us (i.e., the region covering PA from 31$\degr$ to 211$\degr$). The outer molecular gas emission does not appear to be associated with any clear dust structure, at the level of the sensitivity of the Hubble map, which is compatible with a quite lower column density of absorbing material in those areas. It should be also noted that the inner bright clumps of the radio jets lie further outwards than the extent of the molecular gas distribution, as can be seen in Figure \ref{fig:fig5}.

\begin{figure}
\plotone{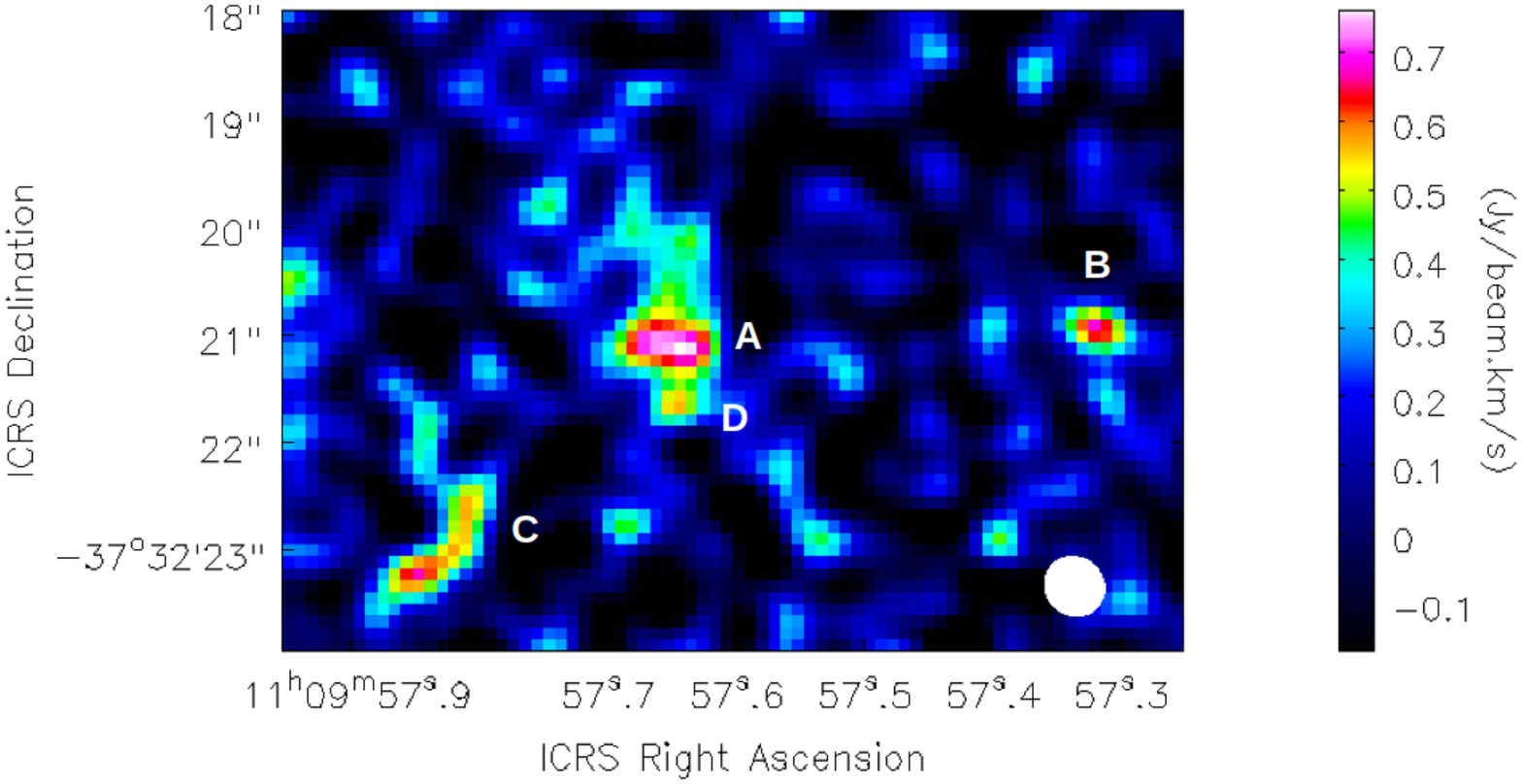}
\caption{CO(1-0) integrated intensity map of the clumps detected in the high-resolution data. Each clump has been identified with a letter code. The synthesized is shown as a filled white ellipse at the lower left.\label{fig:fig6}}
\end{figure}

Using only our higher angular resolution data sets, which results in a synthesized beam of 0$\farcs$58$\times$0$\farcs$53, we identified several clumpy structures in the molecular gas distribution at SNR $>$ 3$\sigma$, whose distribution is shown in Figure \ref{fig:fig6}, and whose properties are listed in Table \ref{tab:clumps}. These clumps do not appear to be directly associated with either of the kinematic "horns" discussed in the next section. Clumps A and D are associated with the brightest central region, while B and C are located in the outer regions. The amounts of molecular gas in these clumps suggest that B and D are comparable to GMCs in the Galaxy, while clumps A and C may be a conglomerate of those, that is, GMAs. This is partially confirmed by measurements of the FWHM of the CO(1-0) lines from these clumps, which give a broad 150-180 km~s$^{-1}$ for clump A (too large to be a single GMC), and narrow lines for clumps B (25 km~s$^{-1}$) and D (20 km~s$^{-1}$), respectively. For clump C, no reliable measurement of the width of the line was possible because of the relatively low SNR of the individual spectral channels (but not of the integrated map). The total amount of molecular in these clumps is about 1/2 the total in NGC~3557, which suggests that the other half of the molecular gas is located in extended structures.

\begin{deluxetable*}{cccccc}
\tablenum{3}
\tablecaption{Clump Properties\label{tab:clumps}}
\tablewidth{0pt}
\tablehead{
\colhead{Component} & \colhead{Center R.A.} & \colhead{Center Declination} & \colhead{Integrated Flux} & \colhead{Total Molecular Mass} \\ 
\colhead{} & \colhead{(hh:mm:ss.s)} & \colhead{(dd:mm:ss.s)} & \colhead{(Jy~km~s$^{-1}$)} & \colhead{(M$_\odot$)}
}
\startdata
A & 11:09:57.6 & -37:32:21.1 & 0.93 & 2.54~10$^7$ \\
B & 11:09:57.3 & -37:32:20.9 & 0.29 & 7.80~10$^6$ \\
C & 11:09:57.8 & -37:32:23.2 & 0.67 & 1.83~10$^7$ \\
D & 11:09:57.6 & -37:32:21.7 & 0.12 & 3.39~10$^6$ \\
\enddata
\tablecomments{The molecular gas mass in this Table is the {\it total} molecular gas mass includung molecular Helium. It was derived by multipliying the molecular hygrogen gass mass by the standard 1.36 factor.}
\end{deluxetable*}

\section{Molecular Gas Kinematics} \label{sec:kin}

One of the characteristics of fast rotators is that they show a clear tendency to have a common axis of rotation for the stellar, ionized gas, atomic gas, and molecular gas components \citep{dav11}. In addition, for the specific case of NGC~3557, we have two additional structures that provide particular orientation axes that can also be compared with the molecular gas kinematics, that is, the radio jet \citep{sch02} and the nuclear dust ring detected with the HST \citep{lau05}. 

\begin{deluxetable*}{cccc}
\tablenum{4}
\tablecaption{Orientation Angles of Different Components in NGC~3557\label{tab:angles}}
\tablewidth{0pt}
\tablehead{
\colhead{Component} & \colhead{PA (N to E)} & \colhead{Comments} & \colhead{Ref} \\ 
\colhead{} & \colhead{(deg.)} & \colhead{} & \colhead{}
}
\startdata
Optical Major Axis & 34 & Photometry & 1 \\
Dust Ring Major Axis & 31 & HST, photometry  & 2 \\
Radio Jet & 78/73 & VLA/ALMA & 3,4 \\
Molecular (CO(1--0) and CO(2--1)) & 51 & ALMA, CO Rotation major axis & 4 \\
Ionized Gas (H$\alpha$) & 0 (inner 6$\arcsec$)--30 ($>$6$\arcsec$) & Photometry & 5 \\
\enddata
\tablecomments{References: (1), (2) \citet{lau05}, (3) \citet{sch02}, (4) This paper, (5) \citet{gou94}}
\end{deluxetable*}

\begin{figure}
\plotone{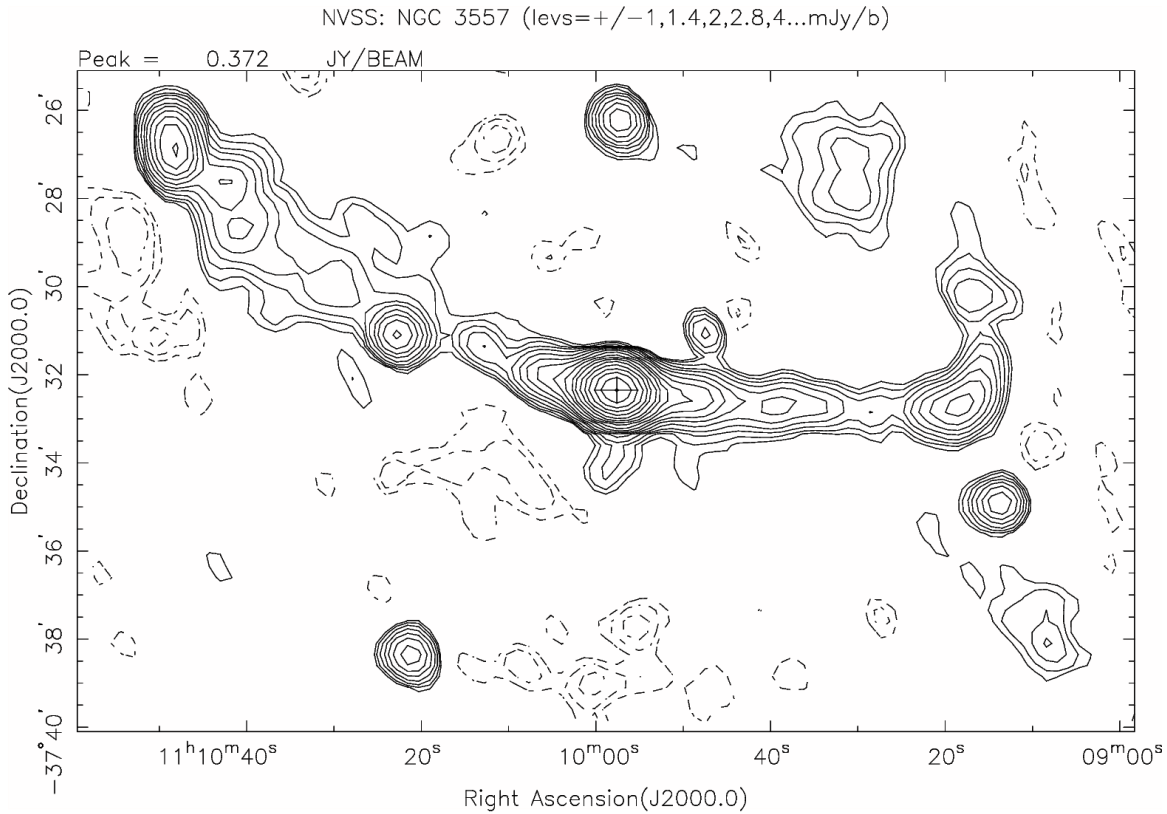}
\caption{Contour map of the NVSS continuum observations of NGC~3557. As indicated in the Figure, the peak flux is 0.372 Jy/beam, and the contours are plotted at $\pm$ 1 mJy/beam multiplied by 2$^0$, 2$^(1/2)$, 2$^1$, 2$^(3/2)$, 2$^2$, ... = 1, 1.414, 2, 2.828, 4, ... .  \label{fig:fig7}}
\end{figure}

We have collected in Table \ref{tab:angles} all the relevant information on the different orientation angles reported in the literature, or determined in this work. Table \ref{tab:angles} clearly shows that the dust absorption appears to be well aligned with the overall shape of the host galaxy. The situation is, however, quite less clear for the other components. The orientation of the inner regions of the radio jet does not appear to be along the minor axis of the overall geometry of the host galaxy, nor perpendicular to the inner dust or molecular gas structures as might be expected from the properties of a fast rotator\footnote{This is well explained by the fact that the SMBH spin axis is not correlated with the large scale molecular disk (e.g.,\citealt{hop12}).}. However, it should be noted that the overall orientation of the jet appears to change at larger scales. As can be seen in Figure \ref{fig:fig7} (data from the NVSS\footnote{NVSS stands for NRAO/VLA Sky Survey.} VLA survey), the jet on the eastern side of the nucleus bends continuously to lower PA reaching at distances of $\approx$10$\arcmin$ an orientation of $\approx$60$\degr$. The counterjet (to the West) appears to have an almost constant PA for about 5$\arcmin$, before turning abruptly North at a very small PA, and possibly turning back on itself. These kinds of abrupt changes are usually ascribed to a combination of jet instabilities/precession and orientation effects \citep{sing16}, and also interactions with the IGM \citep{lan15}. To further investigate the possibility of precession of the radio jet, VLBI imaging is needed, but unfortunately it is not available in the literature so far. It is therefore not possible to ascertain whether the orientation of the jet within the core component of NGC~3557 is oriented closer to the symmemtry axis of the HST dust ring structure. There is also a problen when trying to decide on the orientation of the ionized gas and its kinematics in the inner regions of NGC~3557. The H$\alpha$ photometric image by \citet{gou94} shows that the ionized gas appears to align with the optical photometric axis of NGC 3557 at radial distances $>$ 6$\arcsec$, but the inner regions have isophotes that are basically oriented NS. Furthermore, there is no published ionized gas kinematics map in the literature. Since the ionized gas distribution in the inner 5$\arcsec$ can be severely affected by local extinction from the dust ring, it is not possible to elaborate more on the relationship between ionized gas and the dust ring at this point, and whether the orientation is as expected for a fast rotator.

The situation is better for the molecular gas component. We have analyzed the kinematics of our CO(1-0) data and the publicly available CO(2-1) data using the $^{3D}$BAROLO code \citep{teo15}. This code derives galaxy rotation curves by fitting tilted-ring models, and has modest SNR requirements (i.e., 3) for the emission-line data needed for succesful fits. The regions with emission lines in the cube are detected by the code itself, and selected internally for fitting. The code allows for a selection of parameters (i.e., number of rings, inclination, major axis orientation, center positions, average systematic velocities, etc.) of the tilted rings to be either fixed or left to vary during the fits. The output is computed in two stages, with a second stage using average values for some of the parameters left free in the first stage. Final outputs include plots of the radial dependencies of the fitted parameters, resulting rotation curve PV diagrams, comparison maps between model and actual data in velocity slices, and complete moment maps of the best fits.

\begin{figure*}
\gridline{\fig{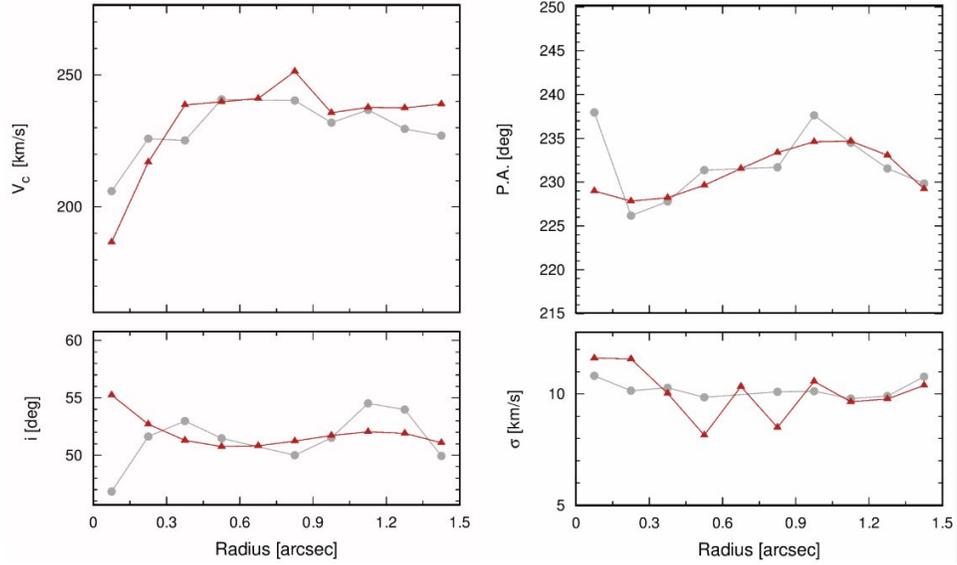}{0.7\textwidth}{(a)}}
\gridline{\fig{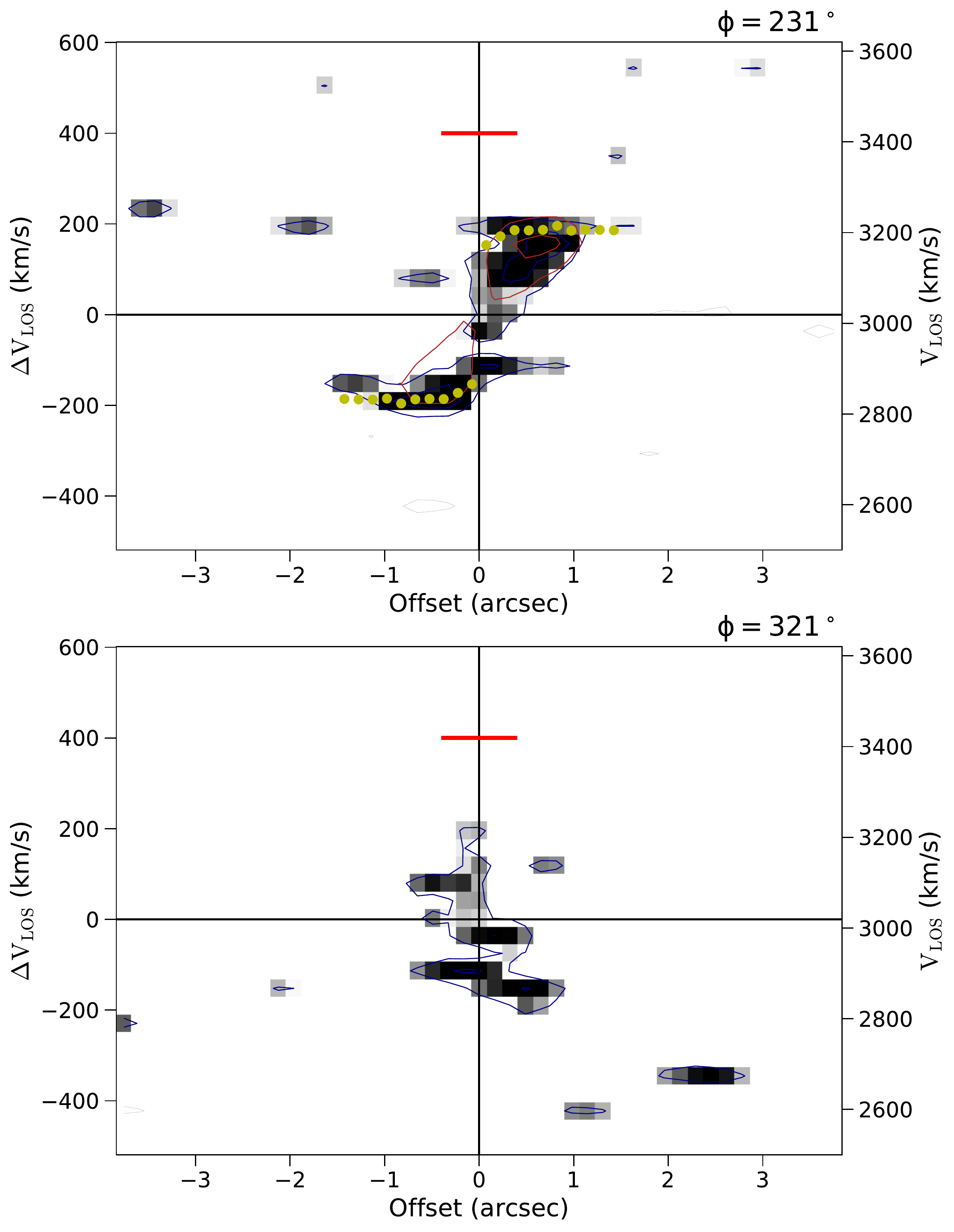}{0.5\textwidth}{(b)}}
\caption{Kinematical properties of the CO(1-0) line emission in NGC~3557 as derived using $^{3D}$BAROLO. (a) shows the final fit results (red curve) for the different annuli (v$_cr$ is the rotation velocity, i the inclination, P.A. the ellipses' major axis PA, and $\sigma_{gas}$ the velocity dispersion, respectively), while (b) shows the fit along the major (upper) and minor (lower) axes superposed on the corresponding PV diagram. The horizontal red line in both plots at 400 km s$^{-1}$ shows the angular resolution achieved by the CO(1-0) observations. See text for additional details.\label{fig:fig8}}
\end{figure*}

\begin{figure}
\gridline{\fig{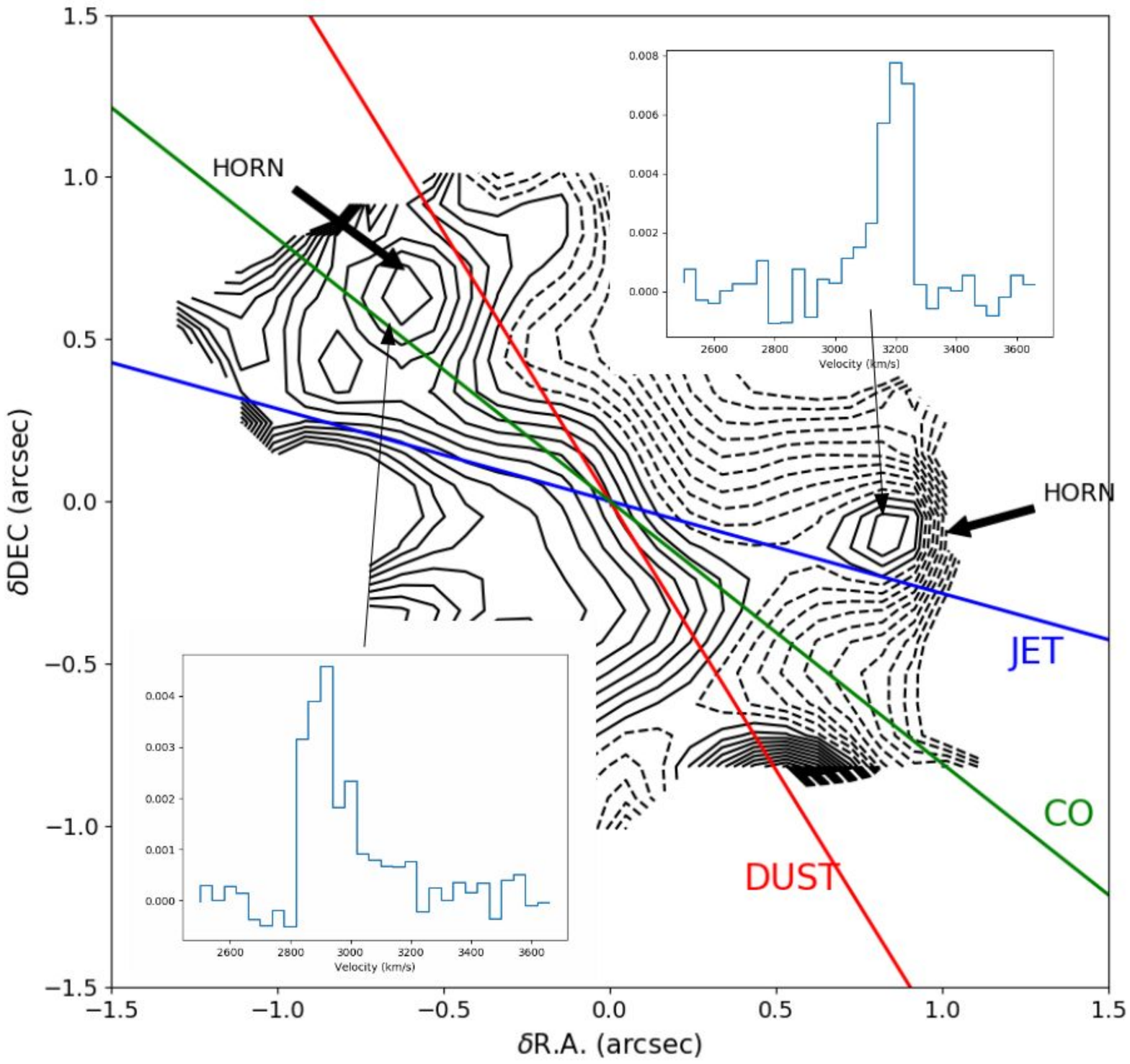}{0.7\textwidth}{(a)}}
\gridline{\fig{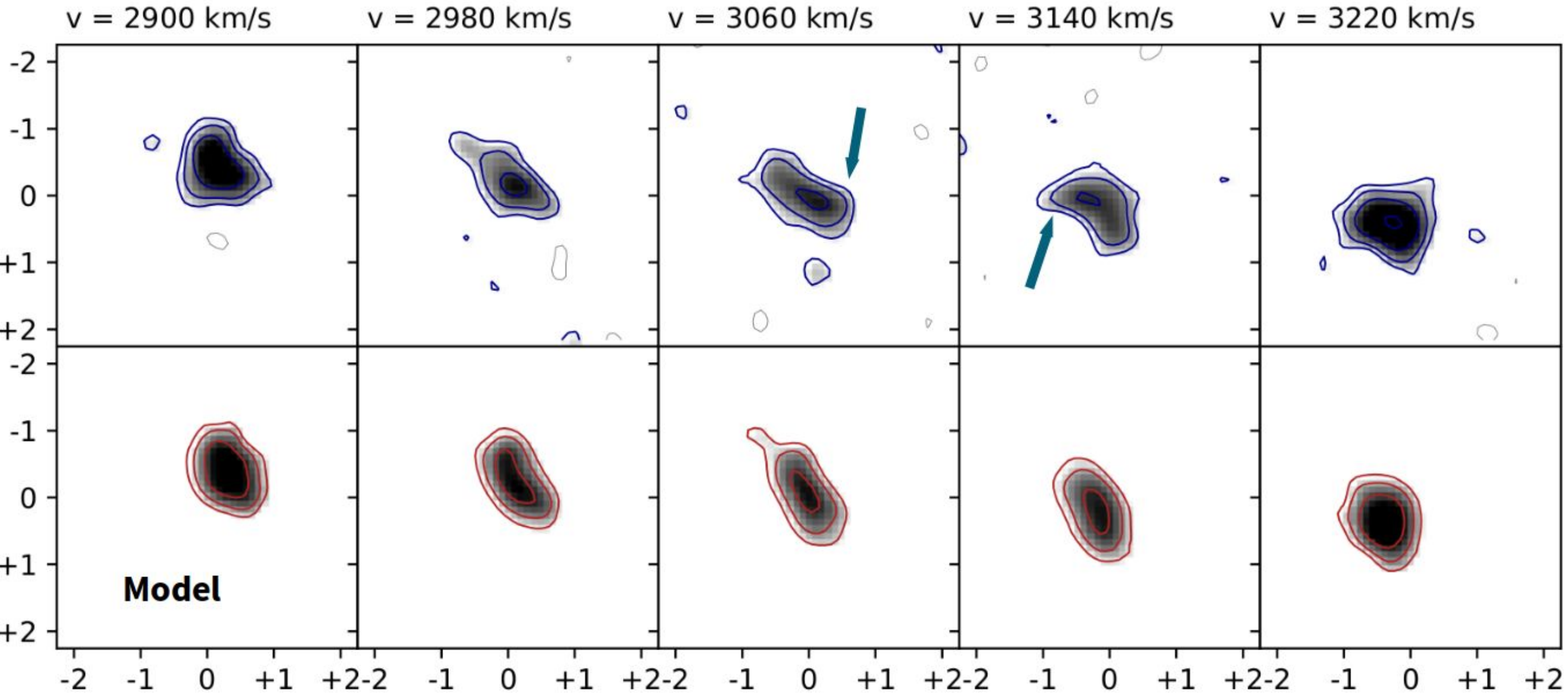}{0.7\textwidth}{(b)}}
\caption{(a) Contour maps of the deviations from pure rotation derived using the difference between the actual data and the rotation model fitted with $^{3D}$BAROLO. The spacing of the contours is 10 km~s$^{-1}$ and negative contours are indicated with dashed lines. Two horn-like structures are clearly seen straddling the nucleus. The radio 3mm jet orientation, the major axis of the HST dust ring absorption, and the kinematical major axis of the CO(2-1) data fits are indicated as blue, red and green straight lines, respectively. The CO(2-1) spectra at the positions of the two horns are also shown. (b) Comparison of the CO(2-1) observations (top) and the $^{3D}$BAROLO fitted kinematical model (bottom) for NGC3557. {\bf The kinematical "horns" are indicated by arrows in the data for velocity slices at 3060 and 3140 km s$^{-1}$, which clearly differ from the fitted pure rotation models shown in the corresponding bottom panels. The scales on both axes are arcseconds}. See text for details.\label{fig:fig9}}
\end{figure}

For both CO datasets we ran the $^{3D}$BAROLO code with a set of initial guesses for the parameters based on our own observations using PV diagrams, and assuming different possible inclinations. The final results are very similar for both CO(1-0) and CO(2-1), suggesting that the fits are reasonably good (the code does not provide error estimates). As can be seen in Figure \ref{fig:fig8}, the average values for the parameters of the rotation curve give a systematic velocity v$_{sys}$ of 3019 km s$^{-1}$, an inclination of 51$\degr$, a PA of the major axis on the receding side of 232$\degr$, and a velocity dispersion of the molecular gas\footnote{The data used for the fits was binned to 40 km s$^{-1}$, and we have checked that the $^{3D}$BAROLO fits converge to similar values for the rest of parameters for initial velocity dispersions in the range 10-40 km s$^{-1}$, as expected.} of $\sigma <$ 40 km s$^{-1}$ (which agrees with the values in other galaxies; e.g., \citealt{com97}). These values are all within 10$\%$ of the values derived for the CO(2-1) data, so we will not repeat those numbers here. Therefore, the data for each CO line independently suggests the presence of bulk rotation of the molecular gas in this object, be it as a bonafide disk or as some kind of inner spirals with significantly symmetric kinematics; the consistency between the two sets of fitted parameters implies further that the model parameters are reasonably accurate. The transition from the linearly increasing to the flat rotation curve appears to happen within 0$\farcs$2 (i.e., 39.6 pc) from the center\footnote{This behavior is seen both in the CO(1-0) and CO(2-1) data, but the actual radial distance at which the transition from steep to flat rotation occurs may be affected by resolution beam smearing effects, since the best angular resolution at our disposal is $\approx$0$\farcs$5 (CO(2-1) datasets).}. This trend is confirmed by the stellar spectroscopy rotation curve along the major axis in \citet{bro07}. This steep rise and flat rotation within tens of parsecs from the center is consistent with the dominance of flat rotation curves in slow-rotating luminous ellipticals reported in \citet{ger01}, although NGC3577 is a fast rotator, as discussed above. The next question is to study the departures from pure rotation, i.e., the non-circular motions, that are unaccounted for with our simple kinematical model. For this purpose, CASA was used to derive the first moment of the velocity field, from which the fitted $^{3D}$BAROLO model was subtracted (see Figure \ref{fig:fig9}). The most interesting result of this is the presence of two kinematic "horns" antisymmetrically placed, w.r.t. the major axis of the dust ring or the host galaxy major axis, straddling the position of the center of the host galaxy. These "horns"  correspond to departures from pure rotation associated with unresolved peaks located antisymmetrically at $\approx$ 0$\farcs$65 (i.e., 129 pc in projected distance) from the center of the host. The peak departures from pure rotation are of +61 km s$^{-1}$ (NE Peak) and +28 km s$^{-1}$ (SW Peak), respectively. Line profiles in the regions of the "horns", show wide low-level velocity components on top of the expected narrower line profile (as shown in Figure \ref{fig:fig9}), consistent with deceleration w.r.t. pure rotation in the SW and acceleration w.r.t. pure rotation in the NE. These wider velocity components also cause that the emissivity-weighted velocity dispersions in the "horns" are larger than for the rest of the molecular gas at similar radial distances, by at least 50 km~s$^{-1}$. At least the "horn" on the NE is also associated with a region of higher gas excitation (see next section).

At the angular resolution of our CO(1-0) observations, the kinematic "horns" lie within the central peak of the continuum emission in NGC~3557 (see Figure \ref{fig:fig5}). It is however possible, given that the ''horns'' and the jet seem to be aligned in projection, that the origin of the kinematic "horns" is related to the interaction of the jet with the molecular gas, as in the case of, for instance, 3C279 \citep{lan15}. 

\section{Disk Stability}
\begin{figure}
\plotone{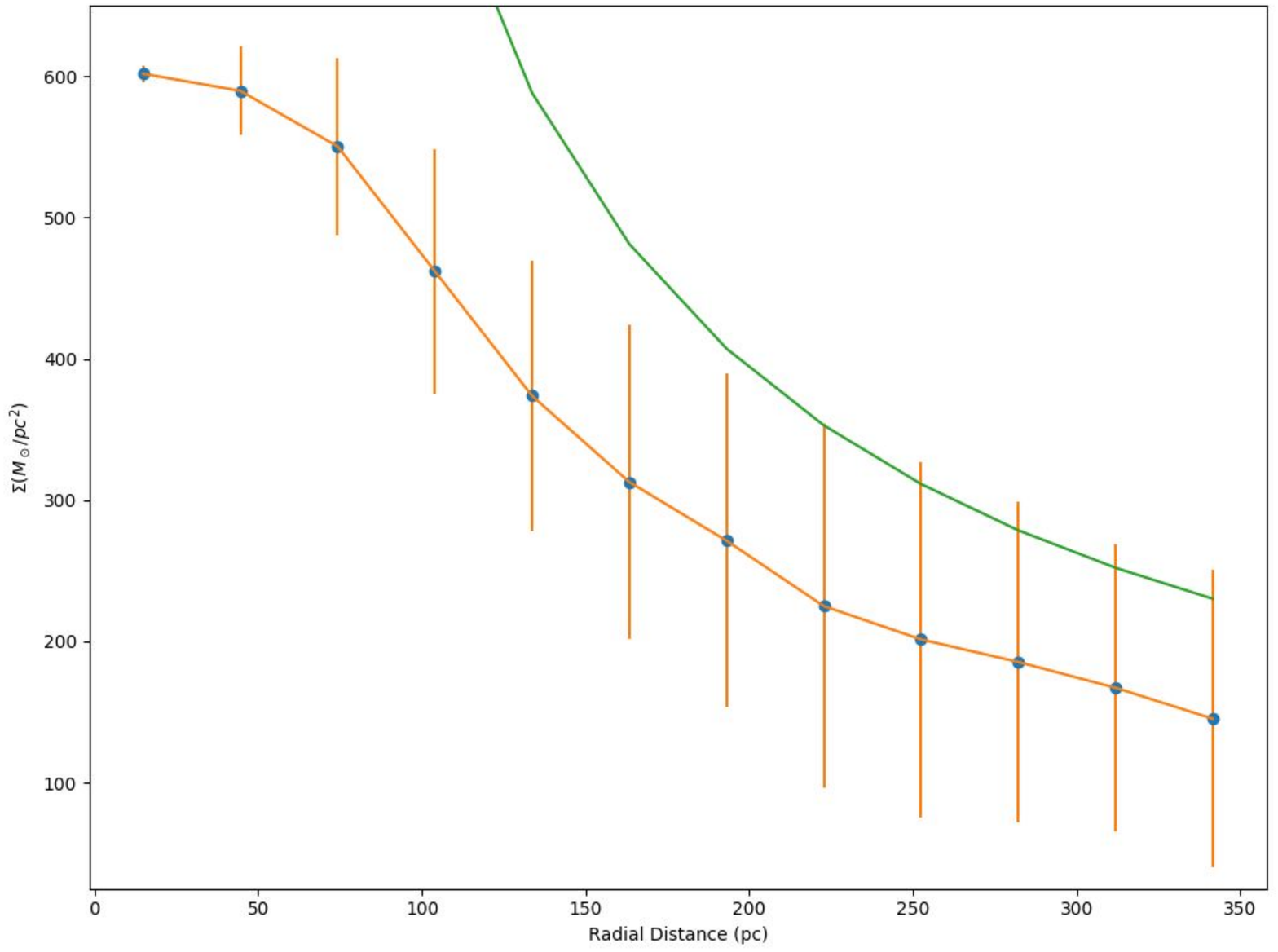}
\caption{Molecular gas surface density as a function of radial distance from the center. The critical surface density curve is also shown in green. Errorbars were derived from the standard deviation of the values along one of the ellipses (see text for additional information). \label{fig:fig10}}
\end{figure}

Assuming that the molecular gas in NGC~3557 is mostly distributed in an inclined disk, and using the rotation curve derived in the previous section, it is possible to study whether the disk can collapse due to local gravitational instabilities and thus form stars. We computed this using the FITS image of the molecular gas integrated intensity (i.e., MOM0) and the inclination of the disk, as derived from the $^{3D}$BAROLO fits, to define a number of concentric ellipses (i.e., inclined rings) separated by a specified spacing. All the pixels in the image that fall onto each of the ellipses were then measured, and the mean, median and standard deviation values per ellipse derived. These values were converted from Jy~km~s$^{-1}$/beam units to intensities (i.e., Jy~km~s$^{-1}$~arcsec$^{-2}$) by dividing by the solid angle of the synthesized beam. The intensity values were then converted to total (i.e., H$_2$ and He) face-on surface mass density of molecular gas using the standard formula \citep{sak95}:

\begin{equation}
\Sigma_{mol}(M_\odot~pc^{-2})~=~6.5\times10^2~cos(i)~I_{CO}(Jy~km~s^{-1}~arcsec^{-2})
\end{equation}

where {\it i} is the disk inclination and $I_{CO}$ the integrated CO(1-0) line intensities. In addition, the software computes the critical surface density for gravitational collapse based on the rotation curve and the criteria from \citet{ken89} as:

\begin{equation}
\Sigma_{crit}(M_\odot~pc^{-2}) = 278~( \frac{\kappa}{km~s^{-1}~pc^{-1}} )~(\frac{\sigma}{6~km~s^{-1}})
\end{equation}

where the epicycle frequency $\kappa$ is derived from the rotation curve as a function of radius as $\kappa=\sqrt{2~V/r~(V/r+dV/dr)}$, and $\sigma$ is the derived velocity dispersion of the molecular gas. The derived surface density and critical surface density are shown in Figure \ref{fig:fig10}. As can be seen in that plot, the surface density decreases rapidly by a factor of almost 6 from the center to a radial distance of 350 pc. Furthermore, the surface density stays clearly {\it below} the critical density for all radial distances, suggesting that local gravitational instabilities are not sufficient to drive star formation in the central regions of NGC~3557 (for other examples of early-type galaxies with sub-critical disks see for instance \citealp{boi17,dav14}). Using the IRAS flux at 24 $\mu$m for this galaxy, and the star formation rate (SFR) calibration in \citep{cal07}, we obtain an upper limit to the SFR\footnote{As stated in \citep{cal07}, in the case of AGNs, the nuclear MIR spectrum can be affected by re-emission of the nuclear torus, etc.} of NGC~3557 of SFR $\leq$0.16 M$_\odot$ yr$^{-1}$. From this and our measured molecular gas mass, the star formation efficiency (SFE) is SFE $\leq$ 1.7~10$^{-9}$ yr$^{-1}$, which, in case the 24 $\mu$m luminosity in NGC~3557 is mostly due to star formation, is comparable with the values reported for spiral galaxies in \citet{ler08}.

\section{Molecular Gas Excitation} \label{sec:ioniz} 

Another important parameter to characterize the properties of the molecular ISM is its excitation. One of the usual methods to derive the excitation of the molecular gas is the use of beam-matched intensity ratios of two rotational transitions of the same species, which, in principle eases at least the issue of possible abundance differences. The ratios are then compared with LTE/non-LTE models to derive average temperatures and particle densities. In the case of elliptical galaxies observed with single-dish telescopes (i.e., moderate angular resolutions), \citet{bvv03} have shown that most objects are sub-thermally excited, with typical CO(2--1)/CO(1--0) values $\approx$0.5 and CO(3--2)/CO(1--0) values $\approx$0.25. There are so far no published works on similar studies, using these two transitions, done at higher angular resolutions on elliptical galaxies. Such studies would provide a resolved map of the molecular line ratios that would yield very important information on the distribution of the excitation within their molecular ISM, possibly hinting to the presence of specific star-formation sites.

\begin{figure}
\plotone{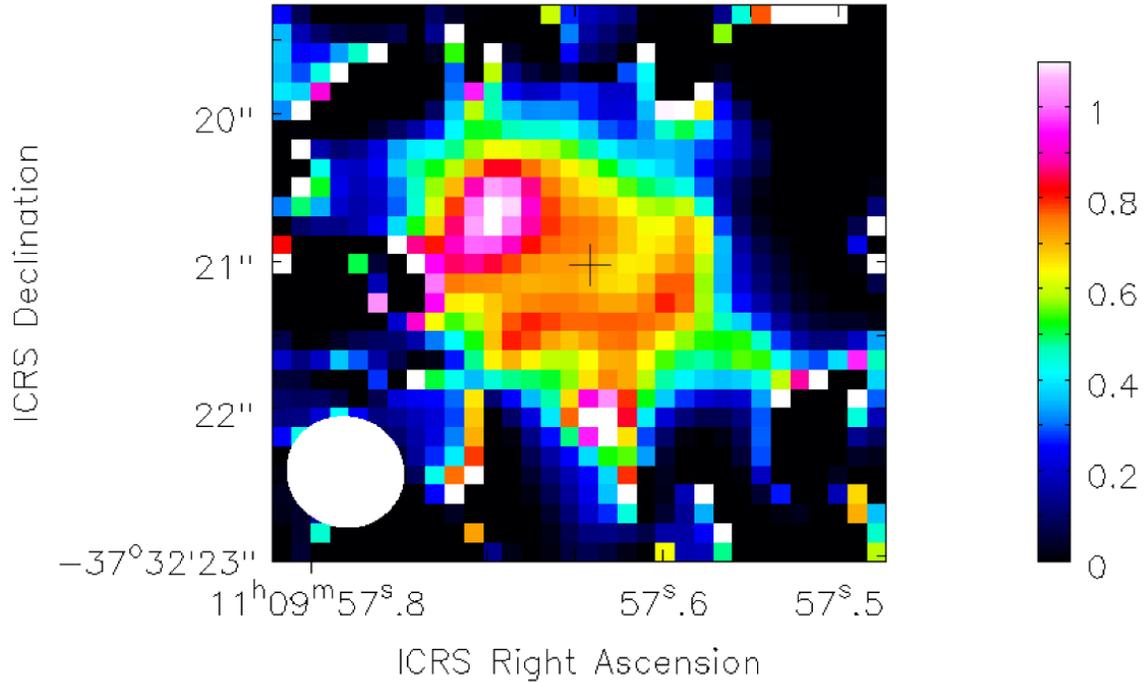}
\caption{Beam-matched ratio of the CO(2-1) and CO(1-0) line intensities (in K~km~s$^{-1}$) of the molecular gas in NGC~3557. The cross indicates the position of the radio continuum peak and the optical center of NGC~3557, and the synthesized beam is shown on the lower left as a filled white ellipse. See text for additional details.\label{fig:fig11}}
\end{figure}

We present here the beam-matched CO(2-1)/CO(1-0) integrated intensity ratio for NGC~3557 created using the moment-0s for both transitions. The original angular resolution of the CO(2--1) data is $\approx$ 0$\farcs$6. As in the case of our data, we subtracted the nuclear continuum emission using the {\it uvcontsub} task in CASA prior to imaging. The images were created with the same pixel sizes as those of our CO(1-0) data (i.e., 0$\farcs$13) and then smoothed with a 2-D Gaussian to obtain the same beam size and orientation as the CO(1-0) data (i.e., 0$\farcs$79 x 0$\farcs$74 along PA 75$\degr$) using the {\it imsmooth} task in CASA, while preserving the flux. The CASA task {\it immath} was then used to create the ratio\footnote{The factor of 4 in the formula is to compensate for the frequency difference between the two transitions. Both maps are in units of Jy km s$^{-1}$.} CO(2-1)/CO(1-0)/4 that is shown in Figure \ref{fig:fig11}.

When comparing emission at two different wavelenghts taken by an interferometer, with possibly different configurations, the issue of the uv-plane sampling is quite important. In our case, the combined data from the two configurations used for the CO(1-0) observations covers uv distances in the range 14.4 - 1100 meters (i.e., 5.5 -- 423.1 k$\lambda$, or, equivalently, angular scales from 117.8 arcsec down to 1.5 arcsec). The CO(2-1) data from the ALMA Archive covers 16.1 -- 438.5 meters (i.e., 12.4 -- 570 k$\lambda$, or equivalently, angular scales from 52.3 arcsec down to 1.13 arsec). The angular scales of the molecular gas structures that we have detected in NGC~3557 are only a few arcsec wide, which implies that the differences in the larger angular scales sampled with the CO(1-0) and CO(2-1) datasets are not significant in deriving the line ratios that we study here.

As can be seen in Figure \ref{fig:fig11}, the ratio shows quite higher values than those obtained in single-dish observations in other ellipticals (see for instance, \citealt{bvv03}). The average value of the ratio in the inner 1.5 arcsec is in fact 0.71$\pm$0.13\footnote{{\bf The errorbars were computed using noise RMS measurements in emission-free areas of the moment 0 images and adding the contribution of a 5$\%$ absolute flux calibration error, as specified by the ALMA project. The final errorbars of the ratio {\it R} are then given as, $\sigma = R~\sqrt{(\frac{\sigma_\alpha}{\alpha})^2+(\frac{\sigma_\beta}{\beta})^2+2(C_{err})^2}$, where $\alpha$ and $\beta$ are the numerator and denominator of the ratio {\it R}, and C$_{err}$ is the nominal 5$\%$ absolute calibration error.}}, and there is a very conspicuos region of higher ratio values (i.e.,$\approx$1.15$\pm$0.35) located 0$\farcs$65 (i.e., 128.7 pc from center in projected distance) to the NE of the center of NGC~3557, as defined by the unresolved radio continuum peak. This region of high ratio values appears to be associated with the area where one of the kinematical "horns" lies (see Section \ref{sec:kin}). Furthermore, it is also associated with a region which shows high velocity dispersions (i.e., 135 km s$^{-1}$ in the CO(2-1) and 270 km s$^{-1}$ in our CO(1-0) data, respectively), which are only surpassed by those at the center in the CO(2-1) data (i.e., 138 km s$^{-1}$ in the CO(2-1) and 180 km s$^{-1}$ in our CO(1-0) data, respectively). In contrast, the region of the counterpart kinematical "horn" in the SW corresponds to a region with ratio $\approx$0.76$\pm$0.24, which is still high when compared with the single-dish values, but is not located in a strong local ratio maximum; instead it lies in some sort of filament or arm of values around 0.7$\pm$0.16 that runs for about 130$\degr$ counterclockwise, starting at PA=135$\degr$. Kinematically, though, this region is also associated to a higher local velocity dispersion peak (i.e., 138.6 km s$^{-1}$ in the CO(1-0) data, compared with neighbouring $\approx$40 km s$^{-1}$). Since we have reasoned in Section \ref{sec:kin} that the horns may be evidence for gas entrainment in the radio jet, it is natural to postulate that jet entrainment may be the responsible for the increase in the value of the ratio, by causing a local increase in density and possibly also in temperature (e.g., \citet{bvv03}), and also explain the larger local velocity dispersions found in both "horns". 
Other possible scenarios for local, high-ratio regions include differences in abundances, local optical depth changes, increased background temperatures, orbit crowding-induced shocks, etc. \citep{izu16}. Since high-energy irradiation by the active nucleus of the host galaxy would be oriented in the same direction as the jet, changes in the local chemistry due to UV and X-ray radiation cannot be ruled out \citep{har10}, but they would have to be accompanied by significant mechanical energy input to explain the larger velocity dispersion observed in these specific areas in NGC~3557. This mechanical energy could come from nuclear shockwaves similar to those found in Cen A \citep{esp17}.

The values of the ratio can be compared with standard single-cloud LVG (code developed by us based on \citet{gold74}) and PDR \citep{kau99} models. Assuming Galactic CO molecule abundances\footnote{log(Z/[dv/dr]) = -4.5).}, we find that a ratio of $\approx$1.1 would need densities n$_{H_2}$$\geq$10$^4$ cm$^{-3}$, and  kinetic temperatures T$_k\geq$40K (LVG) or FUVs of log(G$_o$)$\geq$2 (PDR). Ratio values $\approx$0.7 have more moderate requirements, with densities n$_{H_2}$$\geq$10$^3$ cm$^{-3}$, and kinetic temperatures T$_k$$\geq$10K (LVG) or FUVs of log(G$_o$)$\geq$0 (PDR). The ratio values observed with single-dishes would require even lower densities \citep{bvv03}. For comparison, in our Galaxy, the values of the ratio go from 0.6 in the solar neighbourhood to 0.8--1.0 in the inner 4kpc \citep{han93}. Other nearby spiral galaxies seem to also have a wide range of ratio values for their disks and bulges when observed with single-dish telescopes, with M51 having a ratio of 0.8 for most of the disk and higher at its bulge \citep{gbur93}, and, for instance, NGC4736 having and overall value of 0.5 \citep{ger91}. It is however clear that the central regions show ratio values close to those of the Milky Way (i.e., average of 0.89$\pm$0.06, \citealt{bra93}), and that extremely high values (i.e, $\geq$2) are only found in active star-forming regions in external galaxies (see for instance \citet{loi90}) and in active starbursts \citep{bra93}. Interferometric observations seem to indicate similar trends (see for instance \citet{cas11} and references therein).

NGC~3557 seems therefore not to be following the trends observed in other elliptical galaxies (at lower angular resolutions). Using the rest of targets in our survey, we will try to assess in a separate paper \citep{bvv19} the significance of these results in the context of a sample of bona-fide ellipticals.

\section{Conclusions} \label{sec:conc}

We have presented ALMA interferometric observations of the 3mm continuum and CO(1-0) line emission in the elliptical galaxy NGC~3557. We achieved an angular resolution of $\approx$0$\farcs$75 (or 148 pc at the distance of the host). We detect a molecular gas mass, traced by CO(1-0), of M$_{H_2}$=(9.0$\pm$2.0)x10$^7$ M$_\odot$, which is comparable to other elliptical galaxies. The molecular gas appears to be concentrated within the inner 250 pc in this object, near the location of the central unresolved flat-spectrum source. Our high-sensitivity continuum observations also detect the inner jets associated with the larger scale structures detected at lower frequencies. The kinematical study of the molecular gas suggests that it has some organized rotation with the same orientation than the nuclear dust absorption detected with the HST, and with the overall stellar rotation of the host. Furthermore, if all the molecular gas is contained in an inclined disk following such rotation, the molecular gas surface density is not enough to drive star formation in the central regions of this object. The radio jet does not appear to be oriented perpendicularly to the nuclear dust ring nor along the stellar rotation axis. Given that at larger scales it shows significant bending, it could be precessing. The map of the integrated intensity line ratio CO(2-1)/CO(1-0) indicates that NGC~3557 is peculiar when compared with the average values of this ratio in elliptical galaxies measured with single-dish telescopes. The average value of this ratio appears to be 0.7$\pm$0.2, but there is also a clear peak of excitation (with a value of 1.1) 0.7 arcsec NE of the nucleus. The origin of these high values is not known, but may be related to the interaction of the molecular gas with the radio jet plasma. This region of high excitation is also located in an area of local acceleration of the molecular gas and higher local velocity dispersion ("horns" in this paper), which further supports the scenario of jet interaction (other scenarios are also discussed).

\end{document}